# Modelling risk for commodities in Brazil: An application to live cattle spot and futures prices

Renata G. Alcoforado[1,2], Wilton Bernardino[2], Alfredo D. Egídio dos Reis[1] and José A. C. Santos[3]

[1] ISEG & CEMAPRE, Universidade de Lisboa

[2] Universidade Federal de Pernambuco

[3] Universidade do Algarve

**Abstract**

This study analysed a series of live cattle spot and futures prices from the Boi Gordo Index (BGI) in Brazil. The objective was to develop a model that best portrays this commodity's behaviour to estimate futures prices more accurately. The database created contained 2,010 daily entries in which trade in futures contracts occurred, as well as BGI spot sales in the market, from 1 December 2006 to 30 April 2015. One of the most important reasons why this type of risk needs to be measured is to set loss limits. To identify patterns in price behaviour in order to improve future transactions' results, investors must analyse fluctuations in assets' value for longer periods. Bibliographic research revealed that no other study has conducted a comprehensive analysis of this commodity using this approach. Cattle ranching is big business in Brazil given that in 2017, this sector moved 523.25 billion Brazilian reals (about 130.5 billion United States dollars). In that year, agribusiness contributed 22% of Brazil's total gross domestic product. Using the proposed risk modelling technique, economic agents can make the best decision about which options within these investors' reach produce more effective risk management. The methodology was based on Holt-Winters exponential smoothing algorithm, autoregressive integrated moving average (ARIMA),

ARIMA with exogenous inputs, generalised autoregressive conditionally heteroskedastic and generalised autoregressive moving average (GARMA) models. More specifically, 5 different methods were applied that allowed a comparison of 12 different models as ways to portray and predict the BGI commodity's behaviour. The results show that GARMA with order $c(2,1)$ and without intercept is the best model.

**Keywords**: Risk Analysis, Future Price, Commodity, Value at Risk, Boi Gordo Index (BGI), generalised autoregressive moving average (GARMA).

## 1. Introduction

Cattle ranching is big business in Brazil, given that this sector moved 523.25 billion Brazilian reals (about 130.5 billion United States [US] dollars) in 2017. In that year, agribusiness constituted 22% of this country's total gross domestic product (GDP), while livestock profits made up 31% of agribusiness's contribution to the GDP. Beef exports are important to maintaining Brazil's positive trade balance since this commodity made up 3.2% of all exports in 2017 and grew 9.6% in volume and 13.9% in sales over the previous year (Associação Brasileira das Indústrias Exportadoras de Carne, 2018).

The term 'risk' in this context is associated with the likelihood of unexpected events occurring in business ventures, financial investments or any other situations in which large financial losses may occur. For example, auto insurance companies need to estimate the occurrence of claims in order to price insurance premiums offered in the market. Equity investment must also take into account the potential for sharp declines in local and international markets.

Similarly, stock market risk management's goal is to calculate as accurately as possible capital requirements by minimising potential losses (Balsara, 1992; Ferreiro, 2019). An unrealised loss can result from the decision to keep financial assets in portfolios even when the assets' values have dropped significantly. In this case, investors prefer to maintain their position with a devalued asset as they expect its value will recover, reaching a break-even point (i.e. no loss or purchase price). A question that arises in this situation is how to quantify the value of potential overly high losses (Ferreiro, 2019).

A decrease in share value reduces prospective gains or, conversely, an increase in unrealised losses. When confronted with this situation, investors must choose between two contrasting decisions: liquidate the share to conserve the capital or keep it in hopes of recovering the amount invested (Lee, Cho, Kwon & Sohn, 2020). Unrealised losses may thus be recovered by retaining portfolio assets rather than selling and accepting the consequent realised losses. However, if assets continue to depreciate, the unrealised losses can become quite significant. Successful decisions reflect investors' ability to perceive fluctuations in their position's value and minimise the likelihood of making wrong decisions whether by buying, selling or maintaining their financial position (Nowzohour & Stracca, 2020).

One of the main reasons why this type of risk needs to be measured is to facilitate setting loss limits. To identify patterns in price behaviour and improve future transactions' results, investors must analyse fluctuations in assets' values for longer periods. Investors can thus define the maximum potential loss that they are willing to accept during financial transactions (Lee et al., 2020).

According to Ballescá-Loyo (1999), using value at risk (VaR) to manage an asset portfolio allows investors to position their portfolio properly based on risk aversion and typical price fluctuations. VaR also allows investors to incorporate their expectations in terms of future volatilities, enabling more accurate, improved adjustments to their stop-loss limits.

VaR was conceptualised by Markowitz (1952) and Roy (1952) to optimise economic agents' choices when seeking to minimise the probability of high losses. VaR is obtained by estimating loss distributions and tail quantile values that measure distributions' tail weight.

The present study sought to analyse the behaviour of a specific commodity with a significant impact on Brazil's GDP, namely, the Boi Gordo Index (BGI), in order to help investors who trade futures contracts. To this end, the concept of VaR was applied to examine a price series of BGI futures contracts using different methodologies and compare the respective forecasts. The research's goal was to help economic agents make the best decision about the options within these investors' reach to ensure more effective risk management.

The analysed database contained 2,010 daily entries in which BGI futures and spot sales were traded on the market, from 1 December 2006 to 30 April 2015. Among the 2,010 daily entries under study, 1,994 were used to develop models, and the remaining 16 were kept apart for later comparison between actual data and model predictions produced by the different methodologies. Futures prices were obtained with GrapherOC software, and BGI spot prices were accessed on the Centro de Estudos Avançados em Economia Aplicada – Universidade São Paulo (Centre for Advanced Applied Economic Studies-University of São Paulo) (CEPEA-USP) website (see http://cepea.esalq.usp.br/boi/). The modelling was done using R software, after which the method was selected that best addresses the research objectives and has the highest predictive capacity in terms of BGI futures prices' VaR.

This paper is organised as follows. Section two introduces the BGI with reference to commodities and futures prices, as well as relevant definitions. Section three first describes the data analysis process and then the exponential smoothing algorithms, autoregressive integrated moving average (ARIMA), ARIMA with exogenous inputs (ARIMAX), generalised autoregressive conditionally heteroskedastic (GARCH) and generalised autoregressive moving average (GARMA) models. The final section presents the conclusions.

## 2. BGI and Futures Prices

A commodity is merchandise (i.e. raw material or primary merchandise) subject to a few or even no change processes, which is produced in large quantities and priced according to international supply and demand. This type of goods' quality can vary slightly, but commodities are largely uniform and independent of their provenance so they can be traded on international stock exchanges (Chiarella, Kang, Nikitopoulos & Tô, 2016). Various types of commodities can be found in international markets, such as natural gas, gold, soy or cattle.

Brazil is a major producer of various commodities including the one under analysis in the present study – the BGI – which is a male castrated bovine carcass that has a net weight of 16 *arrobas*[1] or more, with a maximum age of 42 months. Futures contracts for this meat may be traded on the Bolsa de Mercadorias & Futuros (Stock Exchange of Merchandise and Futures) market and other international markets such as the Chicago Mercantile Exchange or the Australian Securities Exchange. Commodities' purchase and sale are made through 'futures' contracts that stipulate the quantities and quality of commodities to be traded. This research focused on estimating these futures contracts' VaR and, more specifically, BGI futures, as described in the next section.

Commodity markets have a major impact on economies worldwide as these markets can produce fluctuations in the price of goods essential to countries' optimal functioning (e.g. food and energy). The trade in this type of merchandise may have an effect on many other goods' prices that would otherwise simply be subject to the laws of supply and demand.

---

[1] One *arroba* is approximately 15 kg.

Various authors have developed models of commodities and futures price behaviour in financial markets.

One frequently used model is the GARMA process, also referred to as the Gegenbauer autoregressive moving-average model, which was proposed by Gray, Zhang and Woodward (1989). This model harnesses the properties of Gegenbauer polynomials' generating function. Woodward, Cheng and Gray (1998) further introduced a $k$-factor extension while conducting research in which a two-factor GARMA model was applied to the Mauna Loa atmospheric CO2 data. The results confirm that the model has a reasonable goodness of fit and produces excellent forecasts.

Boubaker and Boutahar (2011) also used a GARMA approach to model exchange rates' conditional mean. The cited authors estimated a GARMA model with integrated GARCH using a wavelet-based maximum likelihood estimator. Subsequently, Creal, Koopman and Lucas (2012) proposed a class of observation-driven time series models known as generalised autoregressive score models.

In addition, Khalfaoui, Boutahar and Boubaker (2015) examined the links between the crude oil market and stock markets of G-7 countries. The cited researchers used multivariate GARCH models and wavelet analysis to detect a significant volatility spillover between oil and stock markets. Caporin, Ranaldo and Velo (2015) estimated GARMA and seasonal autoregressive fractionally integrated moving average (ARFIMA) (i.e. a special case of the multi-factor GARMA) models to analyse time series for precious metals volumes. These series exhibited periodic-stochastic behaviour and the presence of long-range dependence, so the cited scholars used a multi-factor GARMA model.

Yan, Chan and Peters (2017) developed GARMA and special ARFIMA subfamily models to deal with US commodity futures time series of counts. The cited study confirmed consistently long memory structures in these highly liquid and important financial instruments and, more specifically, revealed systematic forecastable patterns in trading behaviour and liquidity. Yan, Chan and Peters (2017) also report that GARMA models present excellent forecasts. Souhir, Heni and Lotfi (2019) further adopted a generalised long memory GARMA model to estimate the conditional mean of electricity spot price time series. The cited researchers applied VaR, conditional VaR and other models to assess electricity market exposure and concluded that electricity market commodities can be adopted to support diversification and hedging against stock market risks.

Deb (2019), in turn, used moving average of moving average, exponential weighted moving average and GARCH models to conduct parametric analyses of Indian equity mutual funds' downside risk. The cited author applied a VaR-based approach and found that time series presented considerable downside risk during the selected period of 1999–2014. Kumar (2019) also examined financial markets, specifically investigating the volatility transmission from developed markets to Asian emerging markets. The latter cited researcher used a time series from 1996 to 2015 and worked with a heterogeneous autoregressive distributed lag model to study spillover effects.

Similar to previous studies, the present research also applied a statistical approach using a GARMA model to analyse data from 1997 to 2015 and relied on the GARMA model's predictive power. The VaR was extracted and then compared to the real data to see if the proposed approach provides a reliable VaR and accurate forecasts.

## 3. Empirical Research on BGI

The data consisted of BGI futures prices acquired with GrapherOC software and BGI spot prices taken from the CEPEA-USP website. The futures price data referred to the period from 11 February 2005 to 30 December 2014, whereas the spot prices came from trades between 23 July 1997 and 30 April 2015.

When a future contract is due in March and another in April, for instance, both contracts will be available to buy and will have a price months before March. In other words, futures contracts' terms overlap, and a common market strategy is also to operate using futures with short maturities. If a contract's last month is chosen, the value in the last days will converge on the spot value, and, after launching the series, the next contract will diverge again because it will be due in 30 days at the end of the futures contract's term. Thus, the data included the next to last month of each futures contract along with the spot prices for the same dates. The resulting period used in the data modelling was from 1 December 2006 to 30 April 2015.

Figure 1 presents a graph of the daily data under study. The black, blue and red lines represent the futures price, spot price without Funrural and spot price with Funrural, respectively. Funrural is a social security contribution tax in Brazil that is levied on gross revenue from sales of rural products. The graph shows a clear positive trend over time.

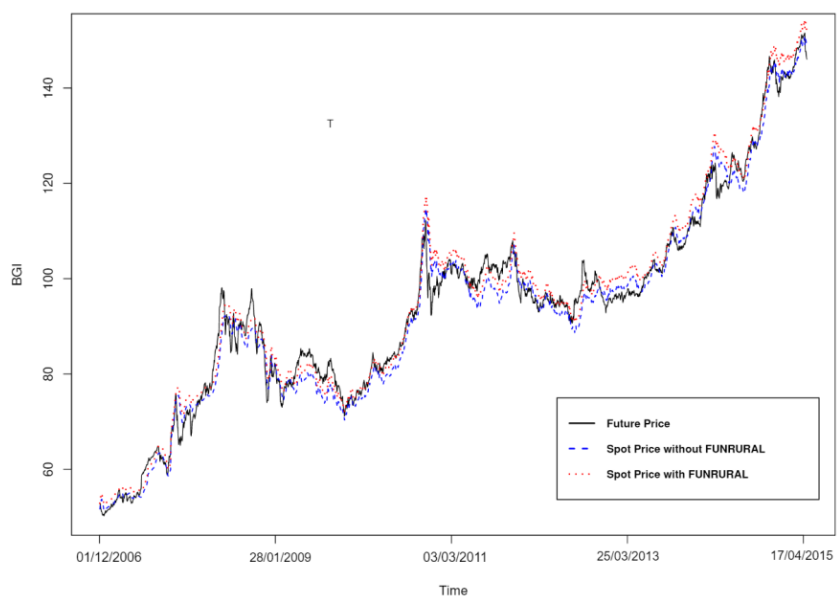

Figure 1: Daily Entries

Figure 2 is a scatter plot of futures prices and spot prices. This graph reveals a strong positive correlation between the two prices.

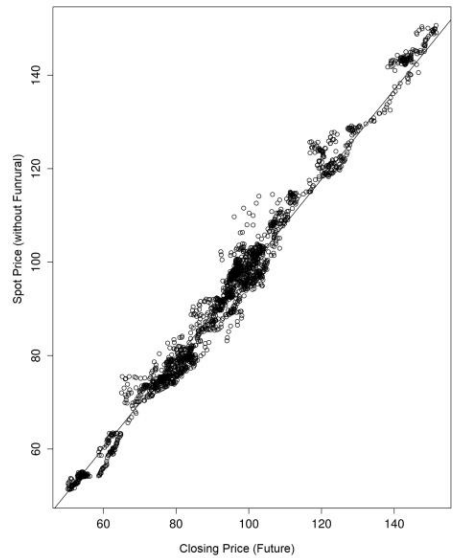

Figure 2: BGI Graph: Futures Price Versus Spot Price (without Funrural)

Figure 3 displays the histogram of BGI futures prices. Most prices are concentrated in the distribution's centre, which at first suggested that they would fit translated $t$-student and normal distributions, although this would entail some positive skewness. However, low prices were a point of concern. This figure shows that the lowest price has a heavier tail, so, when the GARMA models were estimated (see subsection 3.4), a $t$-student distribution was selected as having the best fit.

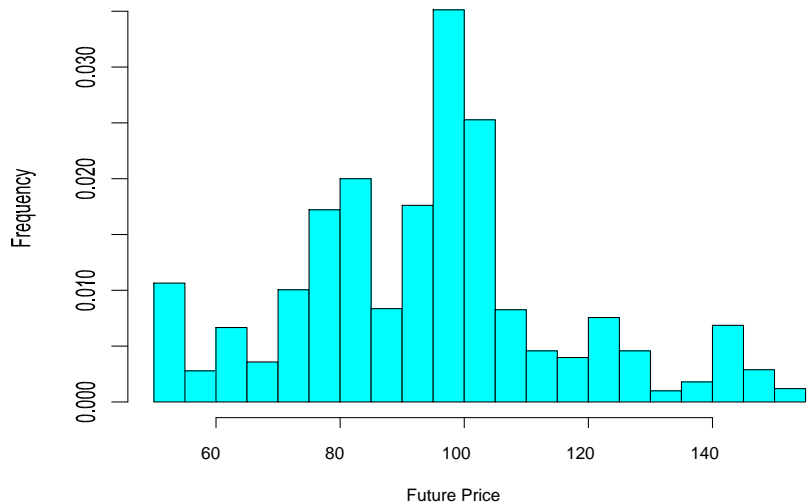

Figure 3: Histogram of BGI Futures Prices

## 3.1 Exponential Smoothing Algorithms

This research relied on simple, Holt and Holt-Winters exponential smoothing models. At first glance, the simple exponential smoothing method resembles the moving average technique in that both approaches extract random behaviour from time-series observations by smoothing historical data (see Brockwell and Davis [2016], chapter 10). The innovation introduced by the simple exponential smoothing model is that this method assigns different weights to each observation in the series. In the moving average technique, the observations used to formulate future value forecasts contribute equally to calculating the predicted values. Simple exponential smoothing, in contrast, uses the latest information by applying a factor that determines its importance (see Makridakis, Wheelwright and Hyndman [1997], chapter 4). According to Wheelwright (1985), the argument in favour of the latter treatment of time series observations is that the latest observations are assumed to contain more information about futures and thus are more relevant to the forecast process.

The simple exponential smoothing model is represented as follows. If $\alpha \in (0,1)$,

then Equations (1) and (2) can be written:

$$N_t = \alpha y_t + \alpha(1-\alpha)y_{t-1} + \alpha(1-\alpha)^2 y_{t-2} + \alpha(1-\alpha)^3 y_{t-3} + \ldots \quad (1)$$

$$N_t = \alpha y_t + (1-\alpha)N_{t-1} \quad (2)$$

in which $N_t$ is the forecast at time $t$ and $\alpha$ is the weight assigned to observation $y_t$. In the present model's fit, $\alpha$ equals 0.9999223, and the forecast coefficient is 150.6. Figure 4 presents the fitted values of the simple exponential smoothing algorithm applied.

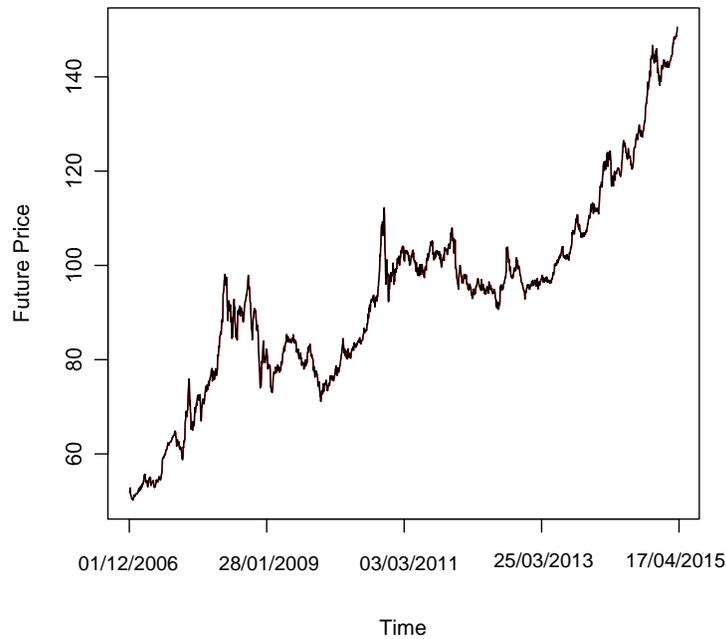

Figure 4: Simple Exponential Smoothing Model

Figure 5 shows a 'zoom in' image of the data, with only the last 31 daily entries. This facilitated a comparison of the real dataset not used in the data modelling with the model's predictions in blue. The dashed blue line is the confidence interval. The mean squared error is 4.982926. This figure highlights the model's forecasting inadequacies.

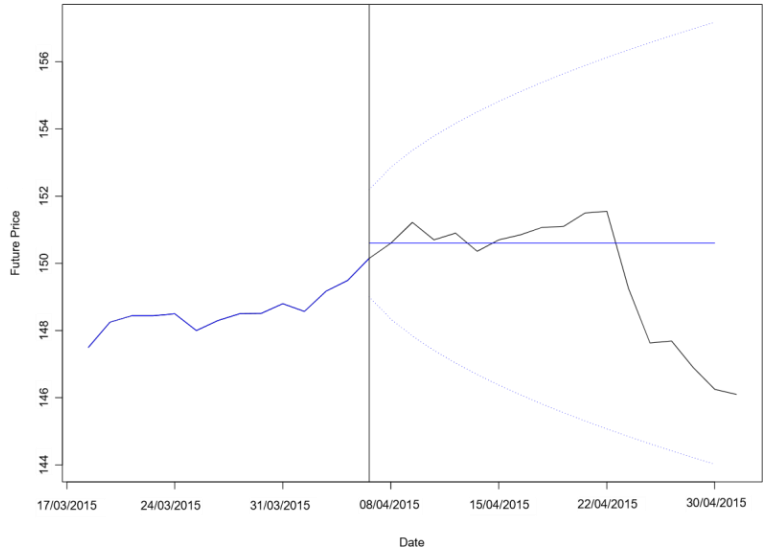

Figure 5: Simple Exponential Smoothing Model's Forecasts

When the simple exponential smoothing method is applied to predicting time series that involve trends in past observations, the predicted values overestimate or underestimate the actual values (Brockwell & Davis, 2016). Thus, the forecasts' accuracy is impaired. The linear exponential smoothing method was developed to avoid this systematic error, in an effort to reflect the trends present in data series (Wheelwright, 1985).

The latter method obtains forecast values by applying Equation (3):

$$\hat{y}_t(h) = N_t + hT_t, h = 1,2,\ldots \qquad (3)$$

in which $N_t$ corresponds to the forecast at time $t$. $T_t$ represents the trend component obtained with Equation (4), and $h$ is the forecast horizon:

$$N_t = \alpha y_t + (1-\alpha)(N_{t-1} + T_{t-1}) T_t = \beta(N_t - N_{t-1}) + (1-\beta)T_{t-1} \qquad (4)$$

in which $\alpha$ is the weight attributed to the observation $y_t, 0 < \alpha < 1$ and $\beta$ is the smoothing coefficient. The present study's Holt exponential smoothing model has the values $\alpha \approx 1^-$ and $\beta = 0.01184473$. The coefficients for the forecast are 150.6 and 0.1450552.

Figure 6 shows, in red, the fitted values for the Holt exponential smoothing model.

This graph looked similar to Figure 4 above, so the next step was to zoom in on the data's last part. Figure 7 shows the results with the simple exponential smoothing model's values in blue and Holt exponential smoothing model's values in red. This graph highlights the differences between the latter two models.

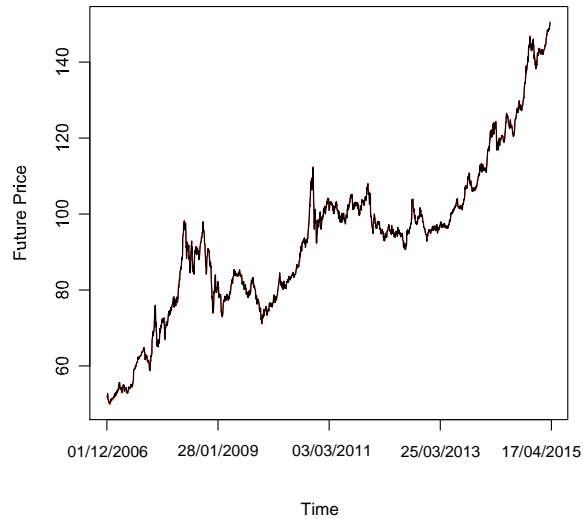

Figure 6: Holt Exponential Smoothing Model

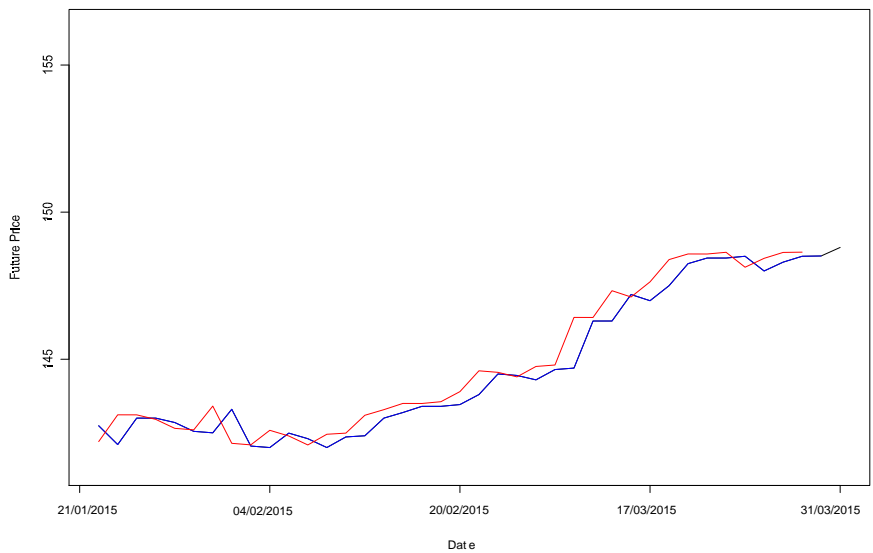

Figure 7: Simple and Holt Exponential Smoothing Models

Figure 8 presents the daily entries from 1980 to 2010 and, in red, the fitted values with the posterior predictions. The dashed lines are the confidence interval, and the mean squared error for the predictions is 14.36051. This figure again shows the Holt model's forecasting inadequacy as the values are clearly overestimated.

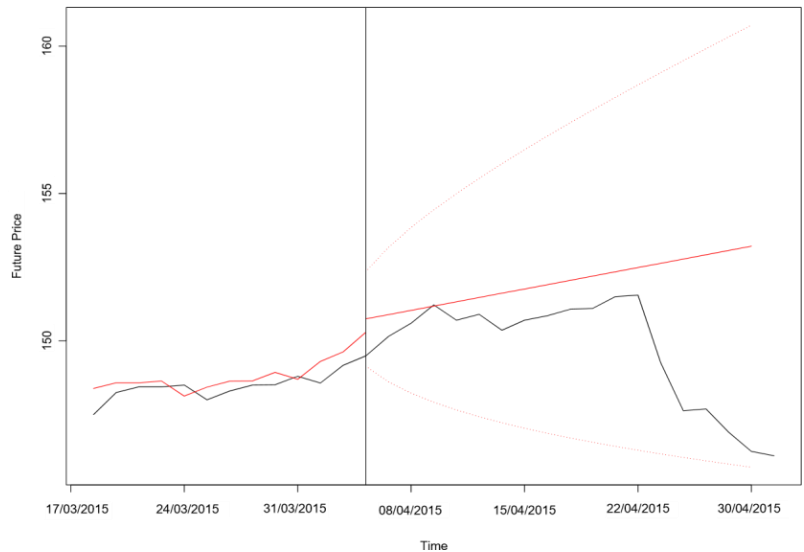

Figure 8: Holt Exponential Smoothing Model's Forecast

The third exponential smoothing algorithm considered was the Holt-Winters approach. Besides the unobservable components present in the previous model, this algorithm also incorporates a third – seasonality (Morettin & de Castro Toloi, 1981). Since the analysed data did not present seasonality, the current study only used the simple and Holt exponential smoothing models. Of these two models, the best fit was the simple exponential smoothing option. Despite only giving fixed value predictions, this model presented a lower mean square error in the forecasts.

## 3.2 ARIMA

This subsection discusses the fit obtained for the futures prices using an ARIMA approach. The ARIMA model was proposed by Box, Jenkins and Reinsel (1970). The present research's first model was the traditional ARIMA(1,1,0) and ARIMA(2,1,1), after which a regression variable was used, namely, an ARIMAX model. A seasonal ARIMA model was not included because, as mentioned previously, the data did not have a seasonal component.

The futures price series were differentiated once, and the autocorrelation function (ACF) was obtained as shown in Figure 9. This graph confirmed that one differentiation was enough, so $d = 1$.

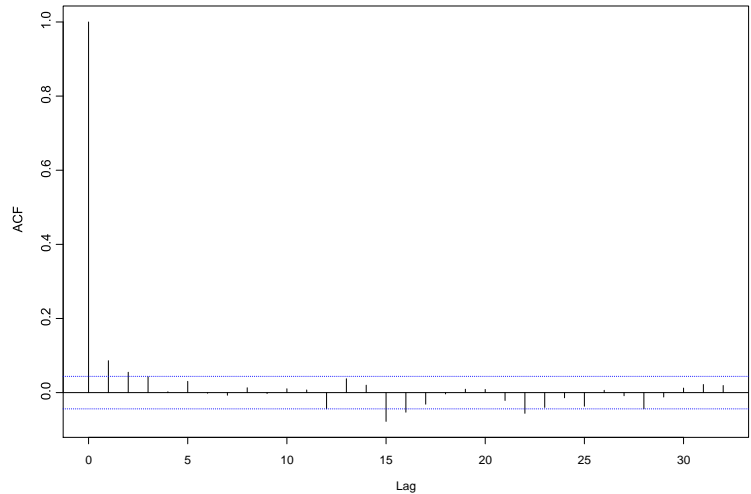

Figure 9: ACF for Futures Prices after One Differentiation

The first line in Figure 9 above represents the ARIMA(1,1,0) results, with an autoregressive coefficient equal to $0.0894$ and an Akaike information criterion (AIC) of $4830.15$. Regarding the diagnostic analysis of this model, the standardised residuals behaved as a white noise sequence, and they did not present either an autoregressive or moving average component.

The second model shown in Figure 9 above is ARIMA(2,1,1), with autoregressive coefficients of $0.6272, 0.0037$ and a moving average coefficient of $-0.5445$. This model has an AIC equal to $4826.42$, which is smaller than the previous model's criterion. Regarding the diagnostic analysis of the ARIMA(2,1,1) model, the standardised residuals again behaved as a white noise sequence, and they did not present either an autoregressive or moving average component.

In both models, the *p*-values for Ljung-Box Q-statistics are higher than 0.05. The futures prices were predicted for 16 days. Figure 10 shows in blue and red the ARIMA(1,1,0) and ARIMA(2,1,1) models' results, respectively. Figure 11 zooms in on only these two models' predictions and presents the futures prices' real value. Neither prediction clearly captured the fall in values happening in the real data. The sum of the predictions' squared errors is 4.761213 and 5.205089 for the ARIMA(1,1,0) and ARIMA(2,1,1) models, respectively.

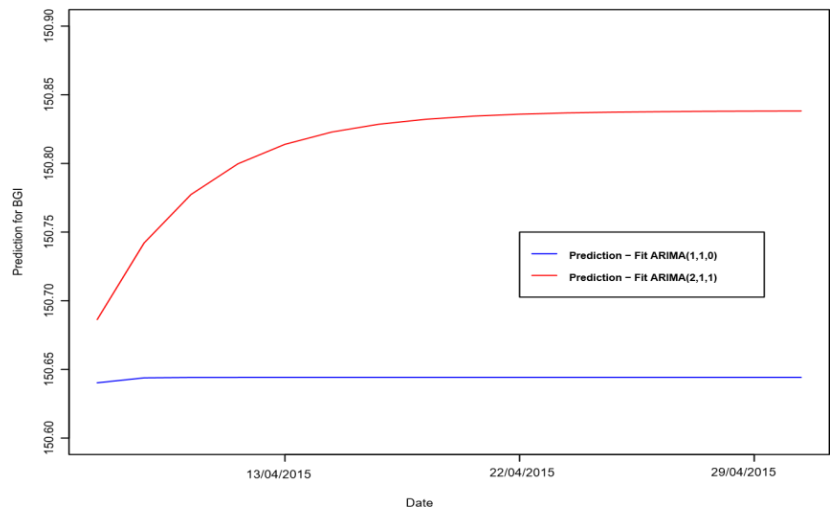

Figure 10: ARIMA(1,1,0) and ARIMA(2,1,1) Predictions

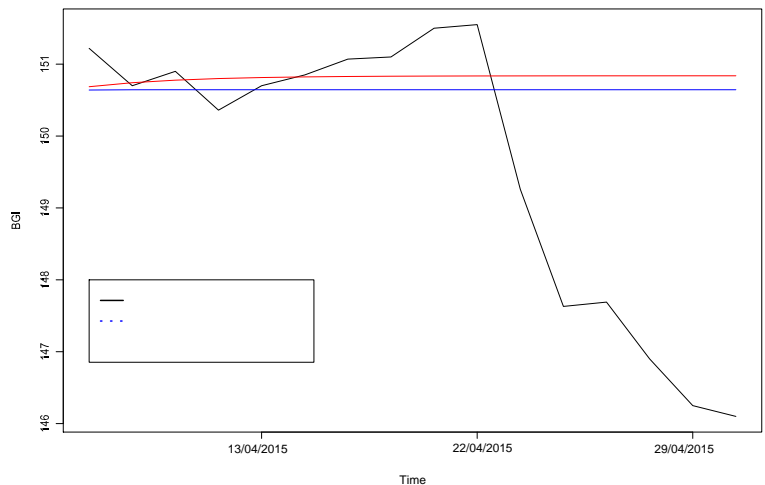

Figure 11: Real Prices and Predictions for ARIMA(1,1,0) and ARIMA(2,1,1) Models

The results support the conclusion that neither prediction and, consequently, neither model was a good fit for this time series. However, more information was available to use in the modelling process, that is, the spot prices. Thus, the next step was to estimate the ARIMAX model with the spot prices as a regression variable. Orders $c(1,1,0)$ and $c(2,1,1)$ were used, as previously, for the purposes of comparison, but this time – after using the auto.arima function in R – order $c(0,0,5)$ was also included.

The ARIMAX(1,1,0) model has an autoregressive coefficient of $0.0048$ and regression coefficient of $0.4416$. The ARIMAX(2,1,1) model produced autoregressive coefficients of $0.0681$ and $0.0116$, a moving average coefficient of $-0.0630$ and a regression coefficient of $0.4388$. The ARIMAX(0,0,5) model's coefficients are $1.1211, 1.1261, 0.9690, 0.6602$ and $0.3271$ for the moving average component, while $3.2594$ was obtained for the intercept and $0.9750$ for the regression.

The models' AICs are 4711.76, 4715.48 and 5606.42, respectively. Figure 12 presents the predictions generated by the ARIMAX models, with sum of squared errors of 6.090724, 6.06713 and 18.64785, respectively. Although the predictions looked somewhat better, the ARIMAX models presented a higher sum of squared errors compared to the ARIMA approach.

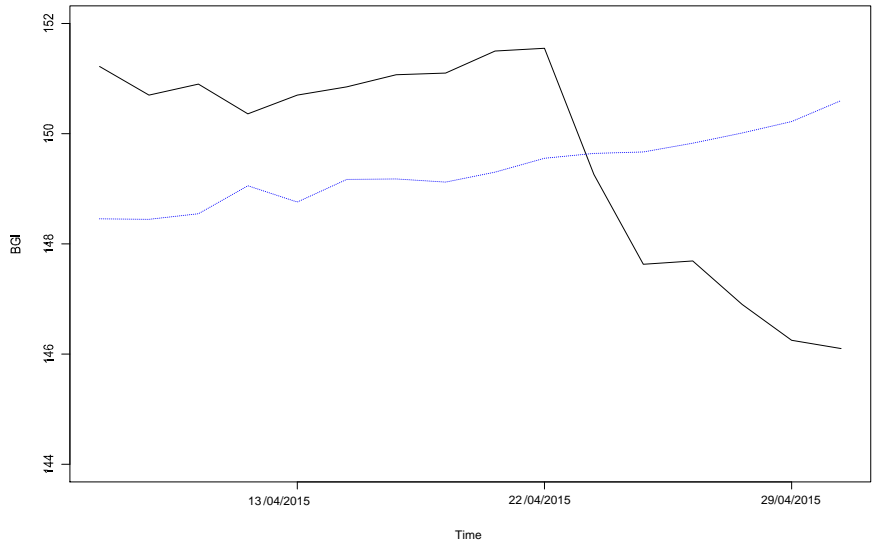

Figure 12: Real Price and Predictions for ARIMAX(1,1,0), ARIMAX(2,1,1) and ARIMAX(0,0,5) Models

Overall, the ARIMA(1,1,0) model presented the lowest sum of squared errors for its predictions, that is, 4.761213. This figure is not only the lowest as opposed to the ARIMA or ARIMAX models' results but also compared with the exponential smoothing algorithms and ARIMAX(0,0,5). The results obtained by using the auto.arima function in R include the highest sum of squared errors for the predictions: 18.64785. Another class of models were estimated to facilitate a retrospective comparison.

## 3.3 GARCH

The GARCH approach was originally proposed by Bollerslev (1986). In the present study, this model was developed using the rugarch package written in R. As described in the previous subsection, the futures prices series were differentiated for modelling purposes, and, after the predictions were generated, the data were transformed back into their original form to compare the forecasts with the real data. Given the dynamic conditional variance, the GARCH model estimated was configured as GARCH(1,1), while the mean model was ARFIMA(1,0,1) and the conditional distribution was normal.

The optimal parameters are $\mu = 0.0538910$, $ar1 = 0.684777$, $ma1 = -0.635453$, $\omega = 0.012223$, $\alpha_1 = 0.080113$ and $\beta_1 = 0.903628$, in which $ar1$ and $ma1$ represent the autoregressive and moving average coefficients. The AIC is 2.2184. Figure 13 shows that the lines are close to each other and the sum of squared errors is $0.6774444$ – the lowest value up to this point.

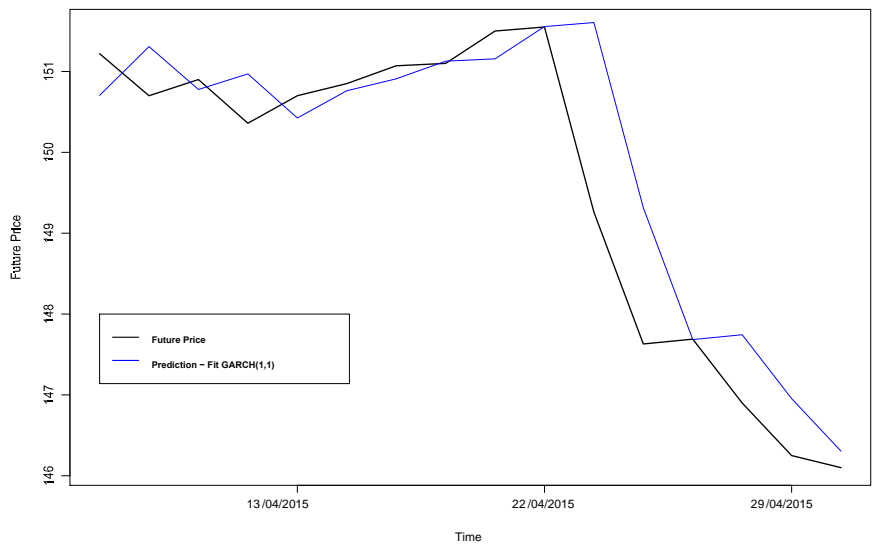

Figure 13: Real Prices and Predictions for GARCH(1,1)

After comparing the GARCH, ARIMA or ARIMAX and exponential smoothing models, the GARCH(1,1) version was found to be the best fit for the data. To finalise the modelling process, one more type of model, GARMA, was estimated for comparison purposes.

## 3.4 GARMA

To apply the GARMA method, the most appropriate family of distributions needed to be determined, as mentioned above in section 3. The *t*-student distribution was selected because the weight of the tails matches the financial data analysed. The orders that best fitted the real figures were $c(1,0)$ and $c(2,1)$. The models with and without intercept were considered to ascertain the best fit.

In this study, the VaR at 5% was always used as it corresponds to the lower limit of a 90% confidence interval in terms of the predictions, that is, a *t*-student distribution quantile. A confidence interval of 90% means 5% in each tail of the distributions, and only the 5% lower weight was of interest in this research. Regarding the distribution, the *t*-student family was used. Stasinopoulos and Rigby (2007) report that the data binding function for $\mu$ should be identity and, for $\sigma$ and $\nu$, logarithmic.

The GARMA model was developed with order $c(1,0)$, that is, a GAR(1) model with intercept. Every estimation has a *p*-value lower than 1%. The resulting model is represented by Equations (5) and (6):

$$\mu_t = \nu_t = 97.011 + 0.3787 x_t + \tau_t \tag{5}$$

with

$$\tau_t = 0.9987 \, A(y_{t-1}, x_{t-1}, 97.011, 0.3787) \tag{6}$$

in which $A$ is the function that represents the autoregressive term, with an associated parameter $\phi = 0.9987$.

The GARMA model with order $c(1,0)$ and without intercept was also a generalised autoregressive GAR(1), except that it had no intercept. All estimates have a *p*-value less than 1%. The model developed is shown in Equations (7) and (8):

$$\mu_t = \nu_t = 0.3961 x_t + \tau_t, \tag{7}$$

with

$$\tau_t = A(y_{t-1}, x_{t-1}, 0.3961), \tag{8}$$

in which $A$ is the function that represents the autoregressive term, with an associated parameter $\phi = 1.00$.

This model was not just autoregressive because it also included a moving average component. The GARMA model used order $c(2,1)$, namely, the autoregressive part had order 2 and the moving average component had order 1. As mentioned previously, all this model's estimations have a $p$-value below 1%. This model is expressed in Equations (9) and (10):

$$\mu_t = \nu_t = 94.5611 + 0.3873 x_t + \tau_t \tag{9}$$

with

$$\tau_t = \sum_{j=1}^{2} A(y_{t-j}, x_{t-j}, 94.5611, 0.3873) + 0.1495\, M(y_{t-j}, \mu_{t-j}) \tag{10}$$

in which $A$ is the function that represents the autoregressive term, with the associated parameter $\underline{\phi}' = 0.8353$ and $0.1631$. $M$ represents the moving average component that has as an associated parameter $\theta = 0.1495$.

These three models' results are presented in Figures 14 and 15. All three are similar, so they are not shown separately. These graphs focus only on the data's last part to be able to show the models by zooming in closely enough. The straight blue line separates the data used for modelling from the part utilised to compare the real data with the models' prediction.

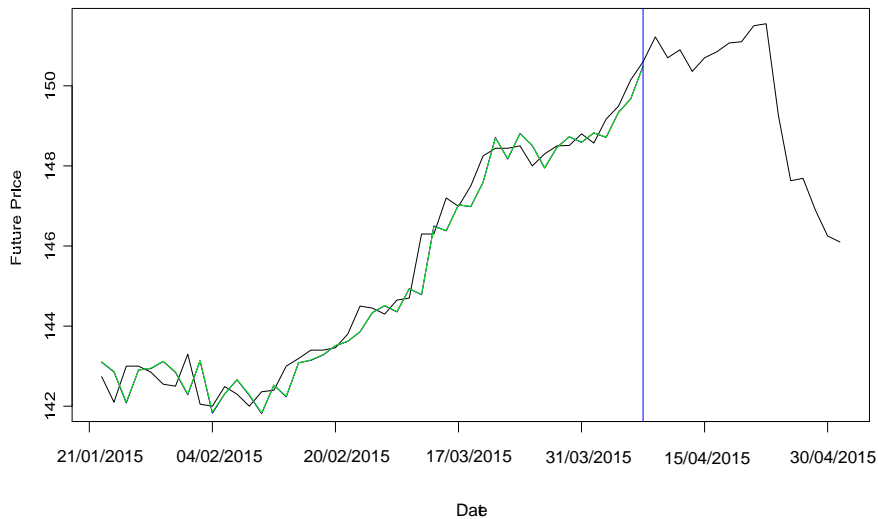

Figure 14: Fit for Last 60 Days of Trade – Order $c(1,0)$ with and without Intercept and Order $c(2,1)$ with Intercept

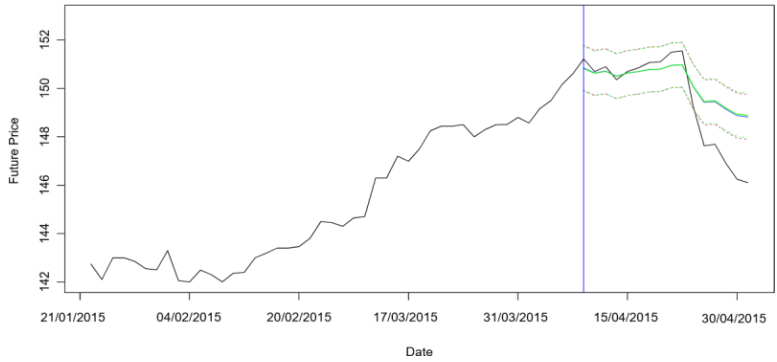

Figure 15: Forecasts of Future Prices with Confidence Interval for April – Three GARMA Models

Figure 15 above presents the models' forecasts. Of the four options analysed, the ones with order $c(1,0)$ with and without intercept and the one with order $c(2,1)$ with intercept also provided similar results in terms of forecasts. The predictions overlap regarding both goodness of fit (see Figure 14 above) and forecasts (see Figure 15 above). In this graph, the three models' futures price values cross the estimated VaR line.

The last model estimated also had an order of $c(2,1)$, without intercept. As in the previous cases, all this model's estimations present a $p$-value lower than 1%. The model is shown in Equations (11) and (12):

$$\mu_t = \nu_t = 1.0086 x_t + \tau_t \qquad (11)$$

with

$$\tau_t = \sum_{j=1}^{2} A(y_{t-j}, x_{t-j}, 1.0086) + 0.1573\, M(y_{t-j}, \mu_{t-j}) \qquad (12)$$

in which $A$ is the function that represents the autoregressive term, with an associated parameter $\underline{\phi}' = 0.7512$ and $0.2013$. In addition, $M$ represents the moving average component with an associated parameter $\theta = 0.1573$.

Figure 16 presents the real data in black, while the blue line marks the separation between data used in the final model and data separated for comparison with the forecasts. The fitted values are in orange. This graph zooms in to show only the last 60 days in order to display the overlapping lines correctly.

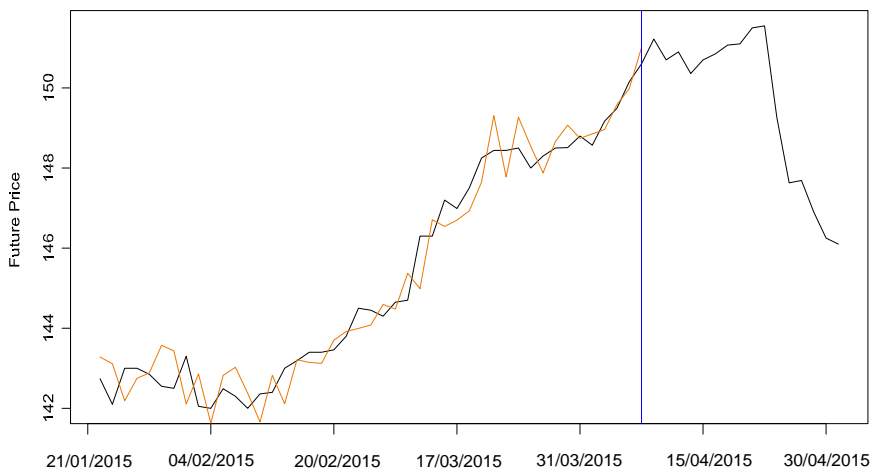

Figure 16: Fit for Last 60 Days of Trade – GARMA Model Order $c(2,1)$ without Intercept

Figure 17 displays the futures price forecasts. This model with order $c(2,1)$ and without intercept stands out as having a better fit. Notably, the graph reveals that the model's forecast curve closely follows the real figures. In addition, the lower confidence interval, that is, the estimated VaR using 95% confidence, is always below the actual future prices at all points, which supports the conclusion that the estimated VaR is reliable. The analyses also included finding the mean square deviation of the four GARMA models estimated. The results are presented in Table 1.

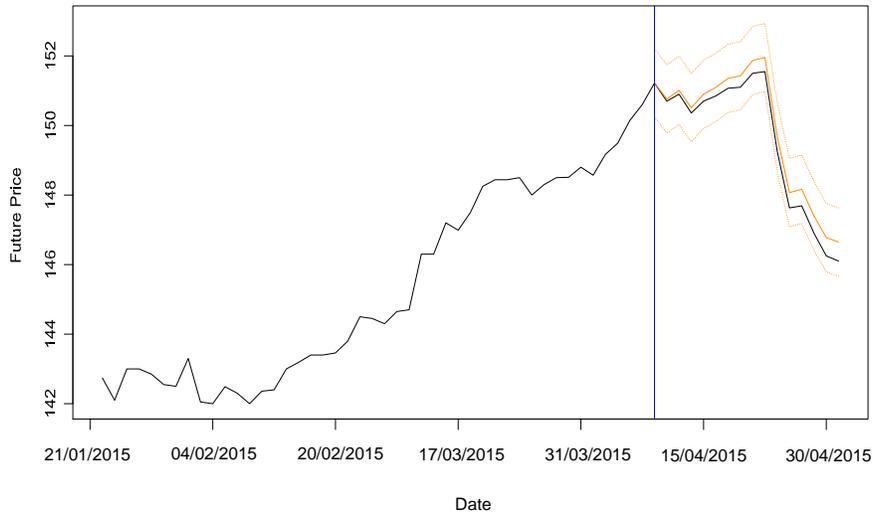

Figure 17: Forecast of Futures Prices with Confidence Interval for April – GARMA Model Order $c(2,1)$ without Intercept

Table 1: Mean Square Deviation of Models' Forecast

| Model | Mean Square Deviation | AIC |
|---|---|---|
| Simple exponential smoothing | 4.9829 | – |
| Holt exponential smoothing | 14.3605 | – |
| ARIMA(1,1,0) | 4.7612 | 4830.15 |
| ARIMA(2,1,1) | 5.2051 | 4826.42 |
| ARIMAX(1,1,0) | 6.0907 | 4711.76 |
| ARIMAX(2,1,1) | 6.0671 | 4715.48 |
| ARIMAX(0,0,5) | 18.6478 | 5606.42 |
| GARCH(1,1) | 0.6774 | 2.2184 |
| GARMA $c(1,0)$ with intercept | 1.7036 | 3965.59 |
| GARMA $c(1,0)$ without intercept | 1.7036 | 3963.59 |
| GARMA $c(2,1)$ with intercept | 1.7862 | 3950.68 |
| **GARMA $c(2,1)$ without intercept** | **0.1284** | 3957.93 |

Table 1 above highlights that the forecast generated by the GARMA model with order $c(2,1)$ is considerably more adequate, and this model's mean square deviation is the smallest. Thus, this model was chosen among the GARMA options as the best way to obtain the VaR of futures prices, as shown in bold in Table 1 above.

The conclusions drawn from the models' fits and Table 1 above are that the two exponential smoothing models did not have as good a fit as the GARMA models and that these two models presented a higher mean square deviation in the forecasts. Therefore, the best model to estimate futures prices' VaR is the GARMA with order $c(2,1)$ and without intercept.

## 4. Conclusions

This study sought to apply the VaR concept in order to analyse the behaviour of BGI futures contract price series using different methodologies so that investors can make better decisions than those offered by the options currently available for effective risk management. Based on empirical research, the model with the best fit in terms of VaR estimations for BGI futures prices is the GARMA with order $c(2,1)$ and without intercept, which was presented in subsection 3.4. This model predicted the BGI futures prices accurately enough that the confidence interval includes the real values.

Since the *t*-student distribution was used, the VaR could be estimated from the quantiles of the $t$ distribution. As shown in subsection 3.4, the model estimated behaved in the desired manner, that is, disposing of all points below the actual futures price values. Thus, this model can protect futures investors by computing the VaR as the maximum loss. In future studies, researchers could examine and analyse different commodities to confirm if the proposed model can also be used to help protect investors against losses in other futures contracts.

## References

Associação Brasileira das Indústrias Exportadoras de Carne. (2018), *Perfil da pecuária no Brasil*, Technical report, Associação Brasileira das Indústrias Exportadoras de Carne, São Paulo, Brazil.

Ballescá-Loyo, L. (1999), "Value at risk and technical analysis", *Technical Analysis of Stocks and Commodities - Magazine Edition.*, Vol. 17, pp. 40–43.

Balsara, N. J. (1992), "Using probability stops in trading", *Technical analysis of stocks commodities*, Vol. 10, No. 5, 225-228.

Bollerslev, T. (1986), "Generalised autoregressive conditional heteroscedasticity", *Journal of Econometrics*, Vol. 31, pp. 307–327.

Boubaker, H. and Boutahar, M. (2011), "A wavelet-based approach for modelling exchange rates", *Statistical Methods and Applications*, Vol. 20, No. 2, pp. 201–220.

Box, G. E. P., Jenkins, G. M., and Reinsel, G. (1970), *Time series analysis forecasting and control*, Holden Day, San Francisco.

Brockwell, P. J. and Davis, R. A. (2016), *Introduction to time series and forecasting*, Springer, Berlin.

Caporin, M., Ranaldo, A., and Velo, G. G. (2015), "Precious metals under the microscope: a high-frequency analysis", *Quantitative Finance*, Vol. 15, No. 5, pp. 743–759.

Chiarella, C., Kang, B., Nikitopoulos, C. S, & Tô, T-D. (2016). The return–volatility relation in commodity futures markets. *Journal of Futures Markets*, 36(2), 127-152.

Creal, D., Koopman, S., and Lucas, A. (2012), "Assessing and valuing the nonlinear structure", *Journal of Applied Econometrics*, Vol. 26, No. 2, pp. 193–212.

Deb, S. G. (2019), "A VaR-based downside risk equity mutual funds in the pre- and post-global financial crisis periods", *Journal of Emerging Market Finance*, Vol. 18, No. 2, pp. 210-236.

Gray, H. L., Zhang, N. F., and Woodward, W. A. (1989), "On generalised fractional processes", *Journal of Time Series Analysis*, Vol. 10, No. 3, pp. 233–257.

Ferreiro, J. O. (2019). Structural change in the link between oil and the European stock market: implications for risk management. *Dependence Modeling*, 7(1), 53-125. https://doi.org/10.1515/demo-2019-0004

Khalfaoui, R., Boutahar, M., and Boubaker, H. (2015), "Analysing volatility spillovers and hedging between oil and stock markets: evidence from wavelet analysis" *Energy Economics*, Vol. 49, pp. 540–549.

Kumar, D. (2019), "Structural breaks in volatility transmission from developed markets to major Asian emerging markets", *Journal of Emerging Market Finance*, Vol. 18, No. 2, pp. 172–209.

Lee, T. K., Cho, J. H., Kwon, D. S., & Sohn, S. Y. (2019). Global stock market investment strategies based on financial network indicators using machine learning techniques. *Expert Systems with Applications*, 117(1), 228-242. https://doi.org/10.1016/j.eswa.2018.09.005

Makridakis, S., Wheelwright, S., and Hyndman, R. (1997), *Forecasting: methods and applications*. Wiley, Hoboken, NJ.

Markowitz, H. (1952), "Portfolio selection", *The Journal of Finance*, Vol. 7, No. 1, pp. 77–91.

Morettin, P. A. and de Castro Toloi, C. M. (1981), *Modelos para previsão de séries temporais*, Instituto de Matemática Pura e Aplicada, Rio de Janeiro, Brazil.

Nowzohour, L. and Stracca, L. (2020), "More than a feeling: confidence, uncertainty, and macroeconomic fluctuations", *Journal of Economic Surveys*, Vol. 34, No. 4, pp. 691–726.

Roy, A. D. (1952), "Safety first and the holding of assets", *Econometrica: Journal of the Econometric Society*, Vol. 20, No. 3, pp. 431–449.

Souhir, B. A., Heni, B., and Lotfi, B. (2019), "Price risk and hedging strategies in Nord

Pool electricity market evidence with sector indexes", *Energy Economics*, Vol. 80, pp. 635–655.

Stasinopoulos, D. M. and Rigby, R. a. (2007), "Generalised additive models for location scale and shape (GAMLSS) in R", *Journal of Statistical Software*, Vol. 23, No. 7, 1-46.

Wheelwright, N. T. (1985), "Fruit-size, gape width, and the diets of fruit-eating birds", *Ecology*, Vol. 66, No. 3, pp. 808–818.

Woodward, W. A., Cheng, Q., and Gray, H. (1998), "A k -Factor garma longmemory model B", *Journal of Time Series Analysis*, Vol. 19, No. 4, pp. 485–504.

Yan, H., Chan, J. S., and Peters, G. W. (2017), "Long memory models for financial time series of counts and evidence of systematic market participant trading behaviour patterns in futures on U.S. treasuries", *SSRN Electronic Journal*, http://dx.doi.org/10.2139/ssrn.2962341